\documentclass[twocolumn,aps,superscriptaddress,showpacs,nofootinbib,floatfix]{revtex4}
\usepackage{epsfig,bm,feynmf}
\usepackage{graphics}
\usepackage{amsmath}
\usepackage[normalem]{ulem}  % \sout{old text} for strikeout
\usepackage[dvips]{color} % For blue in-text comments and additions

\renewcommand\sout{\bgroup \color{red} \ULdepth=-.5ex \ULset}
\begin{document}

%%%%%%%%%%%%%%%%%%%%% Title %%%%%%%%%%%%%%%%%%%%%%

\title{What are the early degrees of freedom in ultra-relativistic nucleus-nucleus collisions?}

%%%%%%%%%%%%%%%%%%%% Authors %%%%%%%%%%%%%%%%%%%%%

\author{P. Moreau}
\affiliation{Frankfurt Institute for Advanced Studies, Johann Wolfgang Goethe Universit\"{a}t, Frankfurt am Main, Germany}

\author{O. Linnyk}
\affiliation{Institut f\"{u}r Theoretische Physik, Universit\"{a}t
Gie\ss en, Germany}

\author{W. Cassing}
\affiliation{Institut f\"{u}r Theoretische Physik, Universit\"{a}t
Gie\ss en, Germany}

\author{E. L. Bratkovskaya}
\affiliation{Frankfurt Institute for Advanced Studies, Johann
Wolfgang Goethe Universit\"{a}t, Frankfurt am Main, Germany}
\affiliation{Institute for Theoretical Physics, Johann Wolfgang
Goethe Universit\"{a}t, Frankfurt am Main, Germany}

%%%%%%%%%%%%%%%%%%%% Abstract %%%%%%%%%%%%%%%%%%%%%

\begin{abstract}
The Parton-Hadron-String-Dynamics (PHSD) transport model is used to
study the impact on the choice of initial degrees of freedom on the
final hadronic and electromagnetic observables in Au+Au collisions
at $\sqrt{s_{NN}}$ = 200 GeV. We find that a non-perturbative system
of massive gluons (scenario I) and a system dominated by quarks and
antiquarks (scenario II) lead to different hadronic observables when
imposing the same initial energy-momentum tensor $T_{\mu \nu}(x)$
just after the passage of the impinging nuclei. In case of the
gluonic initial condition the formation of $s,{\bar s}$ pairs in the
QGP proceeds rather slow such that the anti-strange quarks and
accordingly the $K^+$ mesons do not achieve chemical equilibrium
even in central Au+Au collisions at $\sqrt{s_{NN}}$ = 200 GeV.
Accordingly, the $K^+$ rapidity distribution is suppressed in the
gluonic scenario and in conflict with the data from the BRAHMS
Collaboration. The proton and antiproton rapidity distributions also
disfavor the scenario I.  Furthermore, a clear suppression of direct
photon and dilepton production is found for the pure gluonic initial
conditions which is not so clearly seen in the present photon and
dilepton spectra from Au+Au collisions at $\sqrt{s_{NN}}$ = 200 GeV
due to a large contribution from other channels. It is argued that
dilepton spectra in the invariant mass range 1.2 GeV $< M <$ 3 GeV
will provide a definitive answer once the background from correlated
$D$-meson decays is subtracted experimentally.
\end{abstract}

\pacs{25.75.Nq, 25.75.Ld, 25.75.-q, 24.85.+p, 12.38.Mh}
\keywords{}

\maketitle

\section{Introduction}
\label{sec:intro}

Ultra-relativistic nucleus-nucleus collisions allow to study strongly
interacting QCD matter under extreme conditions in heavy-ion
experiments at the relativistic heavy-ion collider (RHIC) and the
large hadron collider (LHC). The experiments at the RHIC and the LHC
have demonstrated that a stage of partonic matter is produced in these
reactions which is in an approximate equilibrium for a couple of
fm/c~\cite{3years05}. Due to the non-perturbative and
non-equilibrium nature of relativistic nuclear reaction systems, their
theoretical description is based on a variety of effective approaches
ranging from hydrodynamic models with different initial
conditions~\cite{IdealHydro1, IdealHydro2, IdealHydro3, IdealHydro4,
  IdealHydro5, IdealHydro6, ViscousHydro1, ViscousHydro2,
  ViscousHydro3, ViscousHydro4} to various kinetic
approaches~\cite{T1, T2, T3, all, Greco, URQMD, PhysRep, AMPT, BAMPS, PHSD,PHSDrhic}
or different types of hybrid models~\cite{hyb1, hyb2, hyb3, hyb4,
  hyb5,hyb6, hyb7, hyb8}. In the latter hybrid approaches the initial
state models are followed by an ideal or viscous hydro phase which
after hadronic freeze-out is followed up by a hadronic { transport approach
to take care of the final elastic and inelastic hadronic reactions.}

In the ideal or viscous hydro calculations the initial conditions --
at some finite starting time of the order of 0.3 to 0.5 fm/c -- have
to be evaluated either in terms of the (standard) Glauber model or
other initial state scenarios like in the IP-glasma
model~\cite{Schenke,Schenke0}, respectively. Furthermore, a color
glass condensate (CGC)~\cite{GIJV10} is expected to lead {to
structures of smaller scale }  as compared to the Glauber model that
incorporates fluctuations on the nucleon scale. However, in the
hydrodynamic approaches only the equation of state enters as well as
transport coefficients like the shear viscosity $\eta$ that account
for non-viscous phenomena but nothing can be said on the nature of
the microscopic effective degrees of freedom.  This also holds for
hybrid models as long as they employ a hydro phase. To our knowledge
only microscopic transport approaches allow to bridge the gap from
p-p to p-A and A-A collisions in a unique way without introducing
additional (and less controlled) parameters and are sensitive to the
degrees of freedom in the system since their medium dependent
retarded propagators fix the entire time-evolution.

The complexity of heavy-ion collisions is reduced essentially in the
case of proton-nucleus collisions due to the expected dominance of
the initial-state effects over final-state effects. Recently, we
have performed a microscopic transport study of p-Pb collisions at a
nucleon-nucleon center-of-mass energy $\sqrt{s_{NN}}=$ 5.02 TeV and
compared PHSD results to the ALICE measurement at the LHC of the
charged particle pseudorapidity distributions from
Ref.~\cite{ALICE:2012xs} for pseudorapidity $|\eta|<$~2 for
different multiplicity bins of charged particles
$N_{ch}$~\cite{Konchakovski:2014wqa}. However, these differential
pseudorapidity densities did not allow for firm conclusions on the
initial state configuration and the dynamical degrees of freedom
since other approaches compared reasonably well, too: the saturation
models employing coherence effects~\cite{DK12, TV12,
Albacete:2012xq} or the two-component models combining perturbative
QCD processes with soft interactions~\cite{BBG12, XDW12}. On the
other side, a sizeable difference in the mean transverse momentum of
particles $\left<p_T\right>$ versus the pseudorapidity $\eta$ with
opposite slopes in $\eta$ on the projectile side is found within the
CGC framework relative to hydrodynamical or transport
calculations~\cite{Konchakovski:2014wqa}.  Furthermore, an
application of the same approach to Pb+Pb collisions at the
collision energy $\sqrt{s_{NN}}=$~2.76 TeV showed that the heavy
system is not sensitive to the size of initial state fluctuations
\cite{Volo15} when concentrating on hadronic spectra and collective
flow coefficients $v_2(p_T), v_3(p_T)$ and $v_4(p_t)$. This finding
was later on confirmed in a viscous hydro model in Ref. \cite{Miklos15}. In
 these studies the electromagnetic observables had been discarded
 since the degrees of freedom in the initial stage were kept
 unchanged.

In this work we explore the sensitivity of hadronic and
electromagnetic observables to the explicit initial degrees of
freedom under the constraint of an identical energy-momentum tensor
$T_{\mu \nu}(x)$ in the non-equilibrium phase just after the passage
of the two imping nuclei (5\% central Au-Au collisions at
$\sqrt{s_{NN}}=$ 200 GeV). We address this question by investigating
two alternative scenarios for the initial production of the quark
gluon plasma: the scenario I involves purely gluonic initial states
as proposed more than two decades ago in Refs.
\cite{old1,old2,old3,old4,old5} since at that time a first-order
phase transition from hadronic to partonic matter had been expected.
This scenario was recently brought forward again in Ref.
\cite{horst} due to a possibly gluon dominated initial state. The
scenario II describes the initial plasma as a pure ensemble of quark
and antiquark degrees of freedom (without gluons).

After a brief
reminder of the PHSD off-shell transport approach we explain the
implementation of initial gluonic degrees of freedom in Section II
and show the actual dynamical evolution of the quark and gluon
numbers as well as the quark and gluon interaction rates for both
scenarios in case of central Au+Au collisions at $\sqrt{s_{NN}}$ =
200 GeV. The PHSD calculations  for various hadronic spectra,
collective flow  and electromagnetic observables are presented in
Section III in comparison to available data. We summarize our
findings in Sec.~\ref{sec:conclusions}.

%______________________________________________________________________
\section{Reminder of PHSD and its extension}
\label{sec:phsd}

The PHSD model is a covariant dynamical approach for strongly
interacting systems formulated on the basis of Kadanoff-Baym
equations~\cite{JCG04} or off-shell transport equations in phase-space
representation, respectively. In the Kadanoff-Baym theory the field
quanta are described in terms of dressed propagators with complex
selfenergies. Whereas the real part of the selfenergies can be related
to mean-field potentials (of Lorentz scalar, vector or tensor type),
the imaginary parts provide information about the lifetime and/or
reaction rates of time-like particles~\cite{Ca09}. Once the proper
(complex) selfenergies of the degrees of freedom are known, the time
evolution of the system is fully governed by off-shell transport
equations (as described in Refs.~\cite{JCG04, Ca09}). This approach
allows for a simple and transparent interpretation of lattice QCD
results for thermodynamic quantities as well as correlators and leads
to effective strongly interacting partonic quasiparticles with broad
spectral functions. For a review on off-shell transport theory we
refer the reader to Ref.~\cite{Ca09}; model results and their
comparison with experimental observables for heavy-ion collisions from
the lower super-proton-synchrotron (SPS) to
relativistic-heavy-ion-collider (RHIC) energies can be found in
Refs.~\cite{PHSD,PHSDrhic,To12,KB12,Kb12b} including electromagnetic probes
such as $e^+e^-$ or $\mu^+\mu^-$ pairs~\cite{el-m,LiLHC} or real
photons~\cite{photons}.

In the beginning of relativistic heavy-ion collisions color-neutral
strings (described by the FRITIOF LUND model \cite{FRITIOF})
are produced in hard scatterings of nucleons from
the impinging nuclei. These strings are dissolved into
'pre-hadrons' with a formation time of 0.8 fm/c in their rest
frame, except of the 'leading hadrons', i.e. the fastest residues of the
string ends, which can re-interact (practically instantly) with hadrons with a reduced
cross sections in line with quark counting rules.
If, however, the local energy density is larger than the
critical value for the phase transition, which is taken to be $\sim$
0.5 ${\rm GeV/ fm^3}$, the pre-hadrons melt into (colored) effective quarks
and antiquarks as well as massive gluons in their self-generated repulsive
mean-field as defined by the DQPM~\cite{Ca09}.
In the DQPM the  quarks, antiquarks and gluons  are dressed quasiparticles
and have temperature-dependent effective masses and widths
which have been fitted to the lattice thermal
quantities such as energy density, pressure and entropy density.
Furthermore, the interaction rates from the DQPM - entering e.g. in the
electric conductivity or shear viscosity of the hot QGP - have been successfully
confronted with results from lattice QCD (lQCD) \cite{Vitaly,Ca13}.
The nonzero width of the quasiparticles implies the off-shellness of partons, which is taken
into account in the scattering and propagation of partons in the QGP on
the same footing (i.e. propagators and couplings).

The transition from the partonic to hadronic degrees of freedom is
described by covariant transition rates for the fusion of
quark-antiquark pairs to mesonic resonances or three quarks
(antiquarks) to baryonic states, i.e. by the dynamical hadronization
\cite{PHSD,PHSDrhic}. Note that due to the off-shell nature of both
partons and hadrons, the hadronization process obeys all
conservation laws (i.e. four-momentum conservation and flavor
current conservation) in each event, the detailed balance relations
and the increase in the total entropy $S$. In the hadronic phase
PHSD is equivalent to the hadron-strings dynamics (HSD) model
\cite{PhysRep}.

% ____________________________________________________________________
\subsection{Extensions to gluonic initial states}

To modify the PHSD model to gluonic initial states (scenario I)
under the constraint of keeping the energy-momentum tensor $T_{\mu
\nu}(x)$ unmodified with respect to the default PHSD approach we
exchange in the PHSD dissolution routine the massive quark and
antiquark degrees of freedom by massive gluons alone which are
generated by the fusion of flavor neutral quark and antiquark pairs
of closest distance in phase space. Since in PHSD the gluonic
degrees are massive, too, and have broad spectral functions (or
imaginary parts of the retarded propagators), this gluonic fusion
happens at practically the same time as the conventional dissolution
in PHSD (above a critical local energy density of $\epsilon_c
\approx$ 0.5 GeV/fm$^3$ in line with the DQPM and lattice QCD). %%%
We note in passing that the conversion of quarks and antiquarks to
massive gluons is similar to the dissolution in the default PHSD,
however, here the local ratio of quarks and antiquarks to gluons is
fixed by the ratio from the DQPM at the same local energy density
$\epsilon$. Since the DQPM evaluates this ratio in thermal
equilibrium the gluons are substantially suppressed relative to
quarks and antiquarks due to their larger masses, i.e.
$M_g(\epsilon) \approx 3/2 M_q(\epsilon)$. The default PHSD is thus
much closer to the scenario II.
As in the conventional PHSD the gluonic degrees of freedom decay to
quarks and antiquarks in time in line with their decay width and
vice versa ($g \leftrightarrow q + {\bar q}$) and in accordance with
unquenched lattice QCD. In this way energy and momentum as well as
flavor currents are conserved throughout the calculation
\cite{PHSD,PHSDrhic}. The only difference is the 'formation time' of
the 'particles' here which is shorter for the 'gluons' due to their
higher mass and given by the inverse transverse mass $1/\sqrt{{\bf
p}^2+M_g^2}$ in their rest frame where $M_g$ is the mass of the
gluonic quasiparticle in the local cell. In order to introduce the
scenario II in PHSD we immediately decay the formed (colored) gluons
to (colored) quarks + antiquarks according to spectral functions
from the DQPM. We mention that a partly related investigation has
recently been performed in the on-shell transport model of the
Catania group \cite{Grecolast}. We stress, however, that the default
initialization in PHSD includes quark and antiquark as well as
gluonic degrees of freedom that are populated in each local cell
according to the DQPM~\cite{Ca09}. Furthermore, similar strategies have been incorporated in Ref. \cite{Rudy}
for the transition from DQPM degrees of freedom to those from the NJL model 
at a similar initial stage.

\subsection{Parton abundancies}

\begin{figure}[thb]
\centerline{\includegraphics[width=0.4\textwidth]{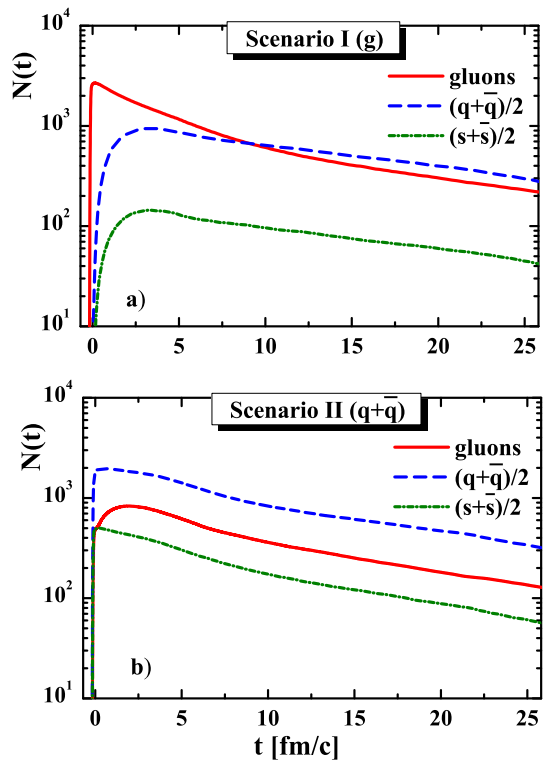}}
\caption{(a) Time evolution of the gluon number (red solid
line) as well as the total quark+antiquark number (divided by 2)
(dashed blue line) and strange+antistrange quark number
(divided by 2) (dot-dashed green line)
as a function of time $t$ for a central ($b$ = 2
fm) Au-Au collision at $\sqrt{s_{NN}}=$ 200 GeV in logarithmic
representation for the scenario I (gluonic initial conditions).
(b) Same quantities as in (a) but for the
scenario II (fermionic initial conditions).} \label{fig1}
\end{figure}

In order to illustrate the procedure we show in Fig. \ref{fig1} the
time evolution of the gluon number (solid red line) as well as the
total quark+antiquark number (divided by 2) (dashed blue line)
and strange+antistrange quark number
(divided by 2) (dot-dashed green line) as a
function of time $t$ for a central ($b$ = 2 fm) Au+Au collision at
$\sqrt{s_{NN}}=$ 200 GeV in logarithmic representation. Initially,
the partonic degrees of freedom in the scenario I are entirely
represented by gluons (from the melting strings) which in time
produce quark-antiquark pairs by gluon splitting. 
Concequently, the strange-antistrange quark pairs appear with a delay.
Since the early
decrease of the gluon number  is approximately exponential we may
attribute a transition time $\tau_g$ to this decay which amounts to
$\tau_g \approx$ 6-8 fm/c which is roughly in line with the BAMPS
calculations from Ref. \cite{BAMPS}. On the other hand the PHSD
calculations with the initial condition of only massive quarks and
antiquarks - as generated by the dissolution of formed hadrons via
string decay in the scenario II - shows a different time evolution
(cf. Fig. \ref{fig1}(b)): here the time evolution exhibits an
approximately exponential decrease of the quark+antiquark number
while the gluons are formed in the first 2-3 fm/c, however, remain
suppressed throughout the time evolution since no chemical
equilibration is achieved.
  \begin{figure}[thb]
\centerline{\includegraphics[width=0.4\textwidth]{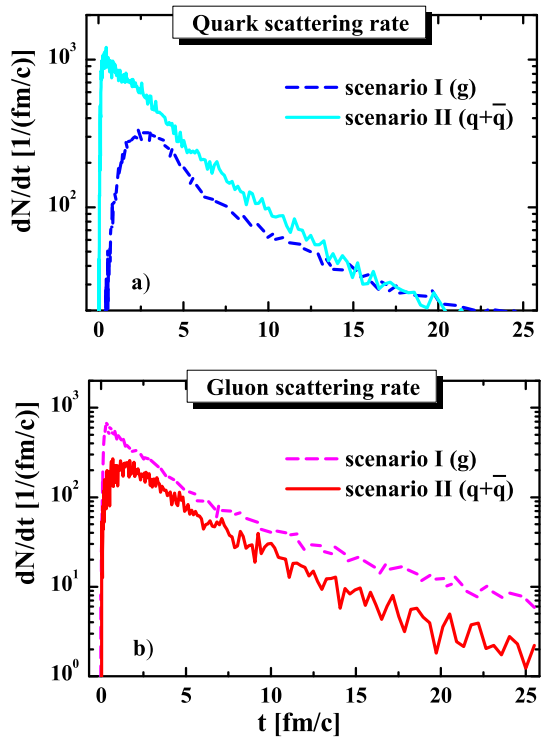}}
\caption{(a) The quark interaction rate $d N_{q}/dt$ from
PHSD as a function of time $t$ for a central ($b$ = 2 fm) Au-Au
collision at $\sqrt{s_{NN}}=$ 200 GeV in the scenario I (lower dark
blue dashed line) and scenario II (light blue upper line). (b) Same
quantities as in (a) but for the gluon interaction rate $d
N_{g}/dt$ (see legend).} \label{fig2}
\end{figure}

It is important to point out that the scenario I is different from
the 'gluonic initial state' proposed in Ref. \cite{horst}. In our
case we use the properties of the retarded propagators for partons
as fixed by the DQPM in comparison with unquenched lattice QCD for
2+1 flavors while in Ref. \cite{horst} a quenched gluonic system is
addressed with undergoes a first-order phase transition at
$T\approx$ 270 MeV and does not couple dynamically to quarks and
antiquarks. Consequently, the degrees of freedom in the model of
Ref. \cite{horst} are color neutral massive glueballs for
temperatures below about 270 MeV whereas in our case we deal with
colored massive gluons in interaction with massive quarks and
antiquarks as inherent in full (unquenched) QCD.

In order to shed some light on the actual dynamics we display the
quark interaction rate $d N_{q}/dt$ (a) and the gluon interaction
rate $d N_{g}/dt$ (b) for both scenarios in Fig. \ref{fig2} for a
central ($b$ = 2 fm) Au-Au collision at $\sqrt{s_{NN}}=$ 200 GeV. In
line with the different initializations the quark interaction rate
is substantially suppressed in the scenario I while the gluon
interaction rate is suppressed in the scenario II. Without explicit
representation we mention that the hadronization rate is slightly
larger in the scenario II than in the gluonic scenario I. The actual
timescales within the PHSD for the transition from gluonic to
partonic matter are  comparable to those from the BAMPS model
\cite{BAMPS}.

\begin{figure}[thb]
\centerline{\includegraphics[width=0.48\textwidth]{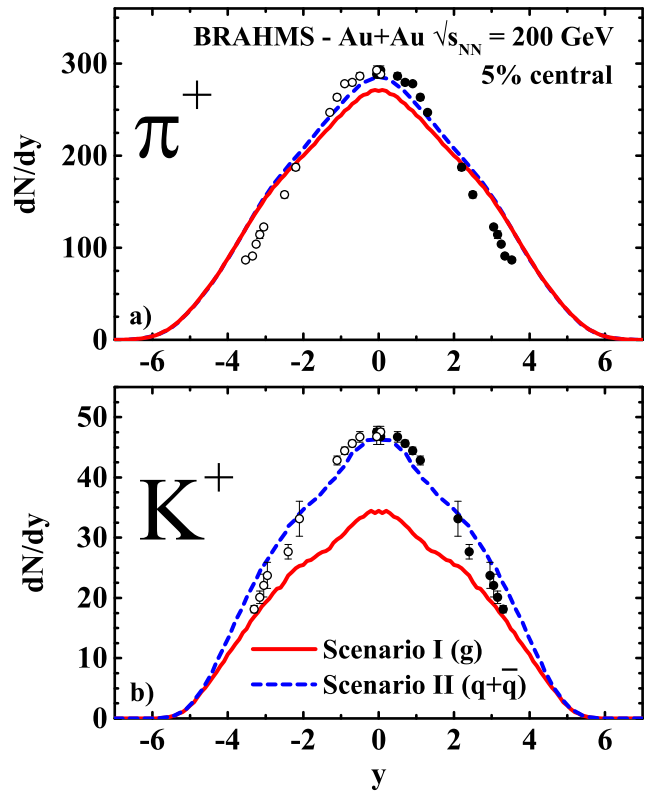}}
\caption{Rapidity distributions for $\pi^+$ (a) and $K^+$ (b) mesons from
PHSD for the scenarios I and II in comparison to the results of the
BRAHMS Collaboration \cite{BRAHMS1} for 5\% central Au-Au collisions
at $\sqrt{s_{NN}}=$ 200 GeV . } \label{fig3a}
\end{figure}

\begin{figure}[thb]
\centerline{\includegraphics[width=0.48\textwidth]{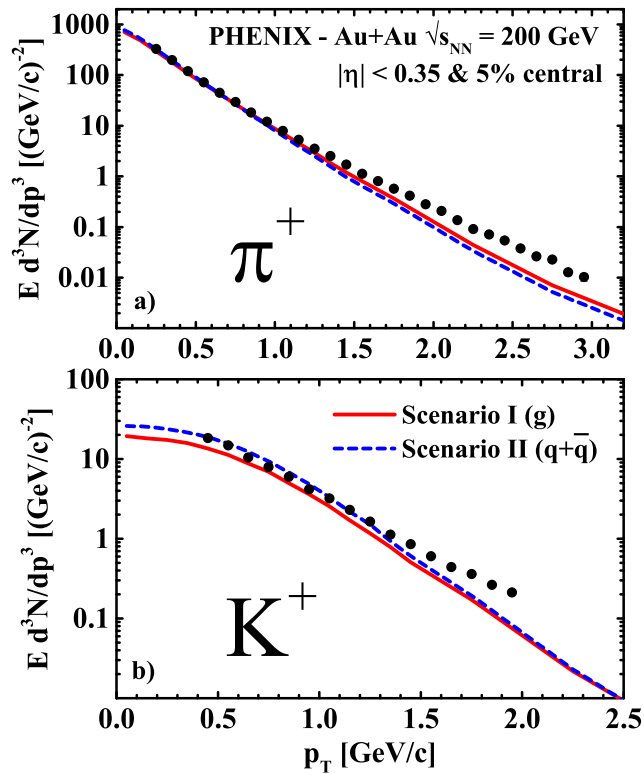}}
\caption{Transverse momentum spectra for $\pi^+$ (a) and $K^+$ (b) mesons
from PHSD for the scenarios I and II in comparison to the results of
the PHENIX Collaboration \cite{STAR2} for 5\% central Au-Au
collisions at $\sqrt{s_{NN}}=$ 200 GeV . } \label{fig3b}
\end{figure}

\section{Comparison to experimental data}
In this Section we show the PHSD results for a variety of
observables from 5\% central Au-Au collisions at $\sqrt{s_{NN}}=$
200 GeV in comparison to experimental data by employing the
different initial state scenarios I (gluons) and II (quarks and
antiquarks). Note, however, that these different scenarios do not
correspond to PHSD although the default version is closer to the
scenario II since in the DQPM the gluons are heavier than the
quarks/antiquarks and suppressed relative to the quarks/antiquarks
for fixed energy density in a given cell.

\subsection{Hadronic observables}
We start with 5\% central Au-Au collisions at  $\sqrt{s_{NN}}=$ 200
GeV and compare in Fig. \ref{fig3a} the rapidity distributions for
$\pi^+$ and $K^+$ mesons from the scenarios I (red lines) and II
(blue dashed lines) with the data from the BRAHMS collaboration
\cite{BRAHMS1}. As expected in the previous Section there is only a
slight difference in the pion rapidity distribution, however, the
$K^+$ distribution is sizeably underestimated in the scenario I
since the strange and antistrange quarks are out-of chemical
equilibrium and in scenario I not present at all in the beginning.
We recall that the strangeness equilibration time for the partonic
energies of interest in the PHSD is in the order of 20-30 fm/c
\cite{Vitalii} which is long compared to the duration of the
partonic phase in central Au+Au collisions at the top RHIC energy.
The slopes of the transverse momentum spectra (cf. Fig. \ref{fig3b})
are quite similar for the two scenarios and slightly underestimate
the data from Ref. \cite{STAR2}.

\begin{figure}[thb]
\includegraphics[width=0.48\textwidth]{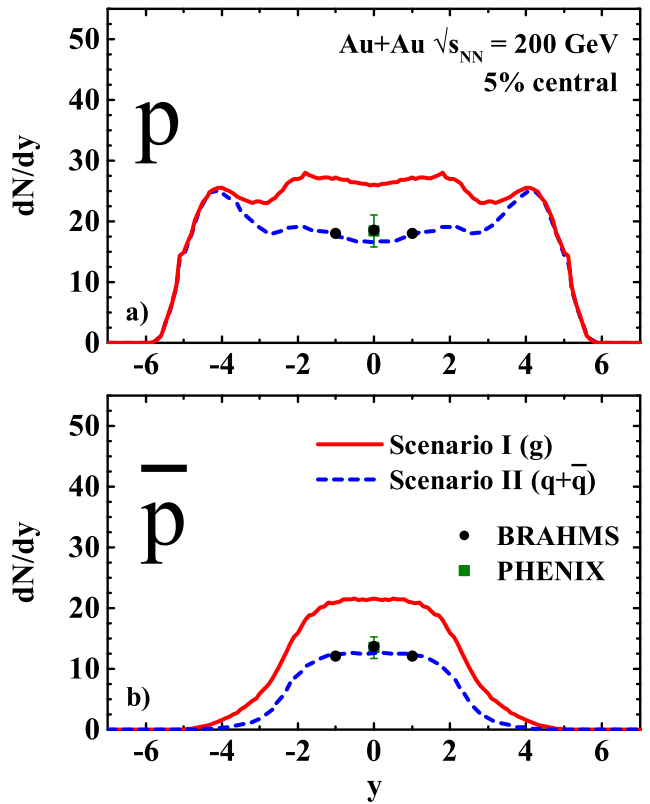}
\caption{The rapidity distribution of protons (a) and antiprotons
(b) for 5\% central Au-Au collisions at $\sqrt{s_{NN}}=$ 200 GeV
from the scenarios I and II in comparison to the experimental data
from the BRAHMS \cite{Arsene:2005mr} and PHENIX collaboration \cite{STAR2}
(the data as well as calculations are without including the
feeddown from strange baryons).} \label{fig4a}
\end{figure}

The results for the proton and antiproton rapidity distributions are
displayed in Fig. \ref{fig4a} for 5\% central Au+Au collisions at
$\sqrt{s_{NN}}=$ 200 GeV in the two scenarios  and demonstrate that the scenario II
with quarks and antiquarks in the initial state is clearly favored
by the data from the BRAHMS \cite{Arsene:2005mr} and PHENIX \cite{STAR2} collaborations.

\begin{figure}[thb]
	\includegraphics[width=0.48\textwidth]{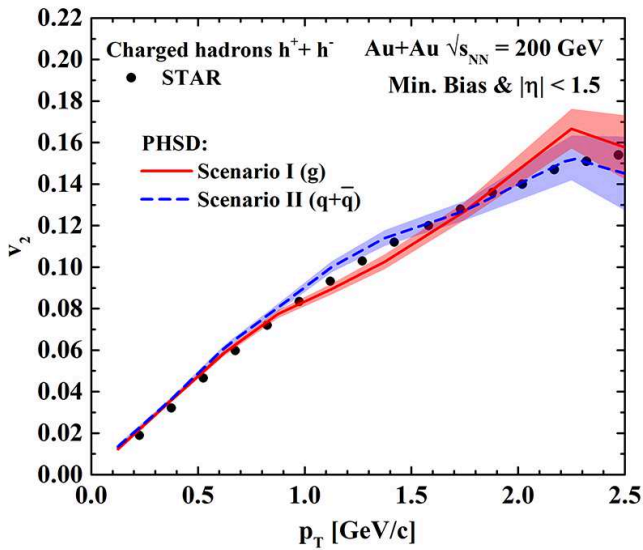}
	\caption{The elliptic flow  $v_2(p_T)$ of charged hadrons for
		Minimum Bias Au-Au collisions at $\sqrt{s_{NN}}=$ 200 GeV from
		the scenarios I and II in comparison to the experimental data from
		the STAR collaboration \cite{STAR3}. } \label{fig4}
\end{figure}

The collective dynamics, however, might show a different picture
since e.g. the elliptic flow $v_2$ is driven by collisions as well
as the repulsive scalar partonic potential as defined by the DQPM.
The actual comparison of
the PHSD results for the scenarios I and II is displayed in Fig.
\ref{fig4} for the elliptic flow $v_2(p_T)$ of charged hadrons for
Minimum Bias Au-Au collisions at  $\sqrt{s_{NN}}=$ 200 GeV.
Unfortunately, there is almost no difference between the two
scenarios while both assumptions are compatible with the STAR data
from Ref. \cite{STAR3}.

\subsection{Electromagnetic observables}

\begin{figure}[thb]
\includegraphics[width=0.45\textwidth]{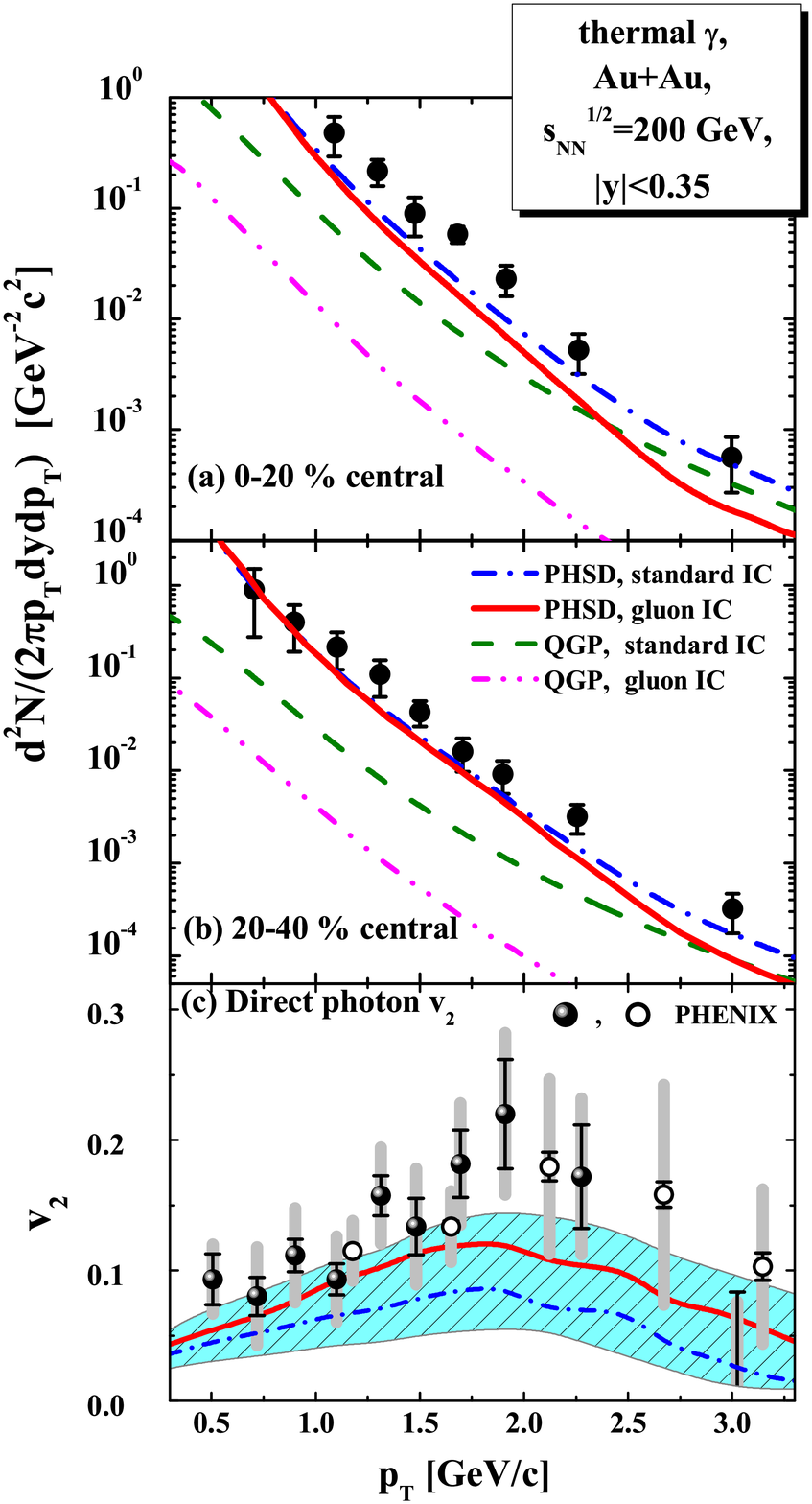}
\caption{The thermal photon yield from partonic channels versus
transverse momentum $p_T$ from PHSD (at midrapidity) in the
scenarios I (lower dash-dot purple lines) and II (dashed green
lines) for  Au-Au reactions at $\sqrt{s_{NN}}=$ 200 GeV for 0-20\%
centrality (a) and 20--40\% centrality (b). The
solid red lines reflect the thermal photon spectrum when adding the
hadronic channels -- dominated by $mm$ and $mB$ bremsstrahlung --
for the scenario I while the dash-dotted blue lines represent the
same quantity for the scenario II. The data for the thermal photons
are from the PHENIX collaboration \cite{PHENIX_thermal}.  The panel c) shows 
a comparison of the elliptic flow $v_2(p_T)$ for the scenario I (solid red line) 
and the scenario II (dashed blue line) with the PHENIX data. 
The hatched area displays the statistical uncertainty of the PHSD calculations.}
\label{fig5}
\end{figure}

We recall that in Ref. \cite{new} the authors have suggested to
investigate asymmetric nucleus-nucleus collisions in order to find
out if in the very initial phase - during the passage time of the
impinging nuclei - electric partonic charges are present since the
asymmetric electric field generated by the spectator protons would
lead to different directed flows of particles and antiparticles (of
opposite electric charge). In the case of symmetric nucleus-nucleus
collisions such an initial  Coulomb boost from the spectator protons
approximately cancels in the center of the partonic medium. In case
of electromagnetic observables we expect the yields and
distributions of photons and dileptons to be quite sensitive to the
initial degrees of freedom (being charged or not). Especially the
production of energetic photons by $q + {\bar q}$ annihilation
should be substantially suppressed in the scenario I \cite{BK2,Trax}
as compared to scenario II where quarks and antiquarks are present
almost from the very beginning. Since also intermediate mass
dileptons were found to be dominated by the $q + {\bar q}$
annihilation \cite{el-m} this expectation should also hold for the
virtual photons.

In order to quantify this expectation we have performed PHSD
calculations for the scenarios I and II following up our previous
studies in Refs. \cite{el-m,photons} where all the details of the
calculations are presented. We recall that also the computation of
the electromagnetic radiation from partons has been evaluated with
the DQPM propagators such that no new parameter enters these
calculations
(cf. Ref. \cite{Rev16} for a recent review).
In Fig. \ref{fig5} we show the corresponding PHSD results for the
two scenarios in comparison to the thermal photon data from PHENIX
\cite{PHENIX_thermal} for 0-20\% centrality (a) and 20--40\%
centrality (b). The partonic photon contribution from $q-{\bar q}$
annihilation and Gluon Compton scattering is displayed in terms of
the green dashed lines for scenario II and the dash-dot purple lines
for the scenario I. The solid red lines reflect the thermal photon
spectrum when adding the hadronic channels -- dominated by $mm$ and
$mB$ bremsstrahlung -- for the scenario I while the dash-dotted blue
line represents the same quantity for the scenario II. We find that
the partonic contribution is about an order of magnitude larger in
the scenario II than in the gluonic scenario I, however, when adding
up all contributions only a moderate depletion of the spectrum is
visible for $p_T >$ 2 GeV/c which is below the PHENIX data at the
largest transverse momenta. Note, however, that the thermal photon
yield is slightly underestimated by the PHSD calculations for the
most central collisions. Accordingly, the present photon data do not
 clearly differentiate between the two scenarios. In addition, the panel c) in  Fig. \ref{fig5} shows
a comparison of the elliptic flow $v_2(p_T)$ for the scenario I (solid red line)
and the scenario II (dashed blue line) with the PHENIX data. Here the 
hatched area displays the statistical uncertainty of the PHSD calculations.
Nevertheless, there is a clear tendency for a larger photon $v_2$ in the scenario I 
which is readily understood in terms of a reduced and delayed production of photons 
in the QGP phase where the photons are emitted from quark and antiquark channels that have 
achieved a finite $v_2$ due to the strong gluon-gluon interactions before. In contrast, the charged hadrons 
are only sensitive to the sum of gluonic and quark/antiquark interactions 
as well as hadronic channels and show no sensitivity to the different 
scenarios in $v_2(p_T)$ as noted before  (cf. Fig. 6). 

\begin{figure}[thb]
\includegraphics[width=0.48\textwidth]{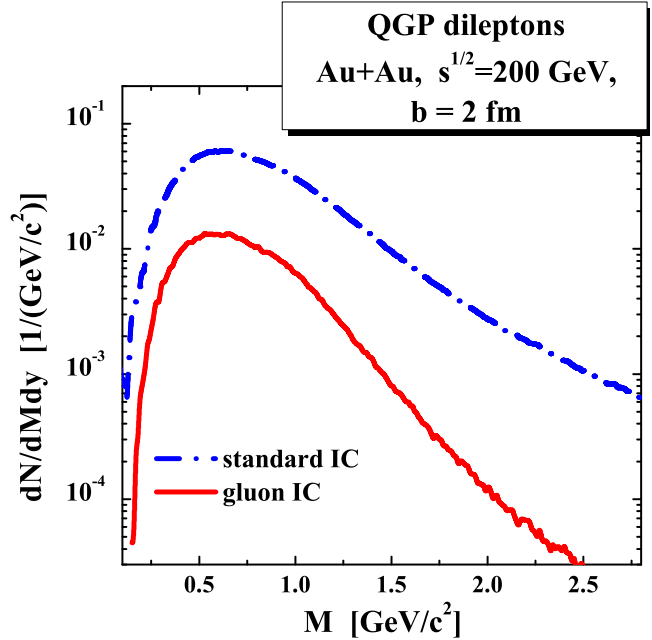}
\caption{Comparison of the QGP di-electron contribution as a
function of the invariant mass $M$ for the scenario I (solid red
line) and the scenario II (dash-dot blue line) for central Au+Au
collisions at $\sqrt{s_{NN}}=$ 200 GeV from the PHSD.} \label{fig6}
\end{figure}

We now turn to dileptons where the virtuality (invariant mass) of
the lepton pair serves as an additional degree of freedom. In  Fig.
\ref{fig6} we show a comparison of the QGP di-electron contribution
as a function of the invariant mass for the scenario I (solid red
line) and the scenario II (dash-dot blue line). Since the dilepton
yield at higher masses ($>$ 1 GeV) is dominated by $q-{\bar q}$
annihilation we find a large difference between the two scenarios
which increases with the invariant mass $M$. Depending on the
background yield and possible subtraction especially the $e^+e^-$
yield from 1.2 GeV $< M < $ 3 GeV should qualify as a proper
observable to distinguish the two scenarios.

\begin{figure}[thb]
\includegraphics[width=0.48\textwidth]{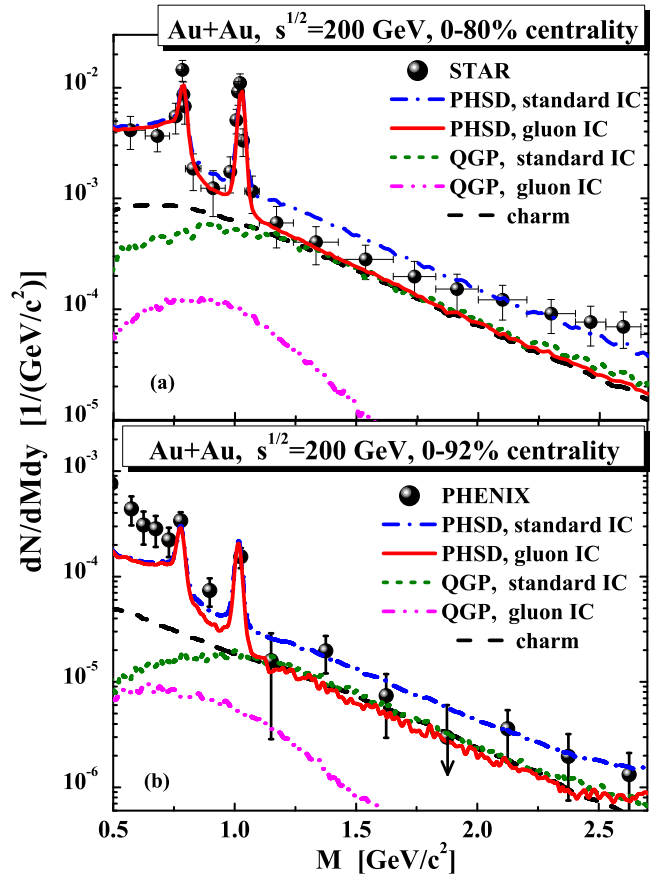}
\caption{The transverse mass spectra of $e^+ e^-$ pairs from  Au-Au
reactions at $\sqrt{s_{NN}}=$ 200 GeV for 0-80\% centrality (a)
and 0-92\% (b). The thermal dilepton yield from
partonic channels is displayed for the scenarios I (lower dash-dot
purple lines) and II (dashed green lines). The solid red lines
reflect the total dilepton spectrum when adding the residual
channels -- dominated by correlated $D$-meson decays -- for the
scenario I while the dash-dotted blue line represents the same
quantity for the scenario II. The dilepton data are from the STAR
Collaboration \cite{STAR_dil} (a) and from the PHENIX
Collaboration  \cite{PHENIX_dil} (b). } \label{fig7}
\end{figure}

In Fig. \ref{fig7} we compare the mass spectra of $e^+ e^-$ pairs
from  Au-Au reactions at $\sqrt{s_{NN}}=$ 200 GeV  for 0-80\%
centrality from STAR \cite{STAR_dil} (a) and 0-92\% from PHENIX
\cite{PHENIX_dil} (b) to the PHSD results for the two scenarios. The
thermal dilepton yield from partonic channels is displayed for the
scenarios I (lower dash-dot purple lines) and II (dashed green
lines) explicitly and shows again  very large differences depending
on the initial degrees of freedom. Whereas in case of scenario II
the QGP contribution is roughly of the same size as the yield from
correlated $D$-meson decays, the QGP yield from the scenario I is
practically not visible in the total spectra. The solid red lines in
Fig. \ref{fig7} reflect the total dilepton spectrum when adding the
residual channels -- dominated by correlated $D$-meson decays -- for
the scenario I while the dash-dotted blue line represents the same
quantity for the scenario II. Whereas the present accuracy of the
PHENIX data does not allow to differentiate the different initial
degrees of freedom, the STAR data  show a better agreement with the
PHSD calculations for the scenario II. However, for a robust
conclusion one needs to subtract the $D$-meson background as e.g. in
the NA60 data \cite{NA60}. This gives a strong motivation for the
STAR detector upgrade with a muon telescope \cite{Ruan:2009ug}.

% ____________________________________________________________________
\section{Conclusions}
\label{sec:conclusions}

In this work the parton-hadron-string dynamics (PHSD) approach has
been employed in the top RHIC energy range for Au-Au collisions in
order to explore the influence of the initial degrees of freedom on
hadronic and electromagnetic observables. For this purpose we have
considered two initial state scenarios: (I)  with only massive
gluons in the initial state (after the passage of the impinging
nuclei) and (II) with only quarks and antiquarks while keeping the
local energy-momentum tensor $T_{\mu \nu}(x)$ unchanged. We point
out that the default PHSD approach does not correspond to these
limiting scenarios, however, is closer to scenario II due to the
thermal suppression of the gluons which are heavier than the
quarks/antiquarks in the DQPM. In PHSD the equilibration between the
gluonic and fermionic degrees of freedom proceeds dominantly via the
channel $g \leftrightarrow q+ {\bar q}$ with an equilibration time
of order 6-8 fm/c. We find that the total partonic collision rates
(adding up quarks/antiquarks and gluons) as well as the
hadronization rate come out not so much different  in the two
scenarios.  However, the formation of $s,{\bar s}$ pairs in the
gluon dominated QGP (scenario I) proceeds rather slow \cite{Vitalii}
such that the anti-strange quarks and accordingly the $K^+$ mesons
do not achieve chemical equilibrium even in central Au+Au collisions
at the top RHIC energy. Accordingly, the $K^+$ rapidity distribution
is suppressed in the scenario I in conflict with the data from
BRAHMS.

Some comments on these results are in order: In the scenario I the
gluon decay to quark-antiquark pairs happens in accord with the
gluon spectral function in PHSD that has a typical width of 100 to
150 MeV which implies that the decay rate is rather 'slow' (on
timescales of 1.5 to 2 fm/c). In the case of gluon decay the ratio
of the strange to light quarks depends on the final phase space and
thus on the quark masses. For the initial times this leads to a
ratio of strange to light quarks of about 1/3, however, the quarks
appear with a delay time of 1.5-2 fm/c relative to the gluons (cf.
Fig. 1a) which is large compared to the average formation time of
partons of 0.2-0.3 fm/c in the scenario II. We recall that the
concept of string melting in the QGP also implies a ratio of the 
strange to light quarks of $\sim$ 1/3 \cite{CSR}.  One might
claim that these widths might be substantially underestimated in the
DQPM (or PHSD) but for substantially larger widths the ratio of the
shear viscosity over entropy density $\eta/s$ or the electric
conductivity $\sigma$ would drop by the same factors and no longer
be in accord with results from lattice QCD (cf. the review
\cite{Rev16}).

The rapidity distributions for protons and antiprotons are
overestimated in the gluonic scenario I while are good agreement is
achieved within the scenario II in comparison to the data from the
BRAHMS and STAR collaborations. The differential elliptic flow of
charged particles  (cf. Fig. \ref{fig4}) is not sensitive to the
initial degrees of freedom, however, a drastic difference is seen in
the photon and dilepton production from partonic sources since the
initial gluonic degrees of freedom carry no electric charge. The
actual comparison to the data from the PHENIX and STAR
collaborations (cf. Figs. \ref{fig5} and \ref{fig7})  slightly favor
the scenario II, i.e. the early presence of quarks and antiquarks,
however, a robust conclusion (from the photon data) will require
more accurate measurements for transverse momenta of 2--3 GeV/c as
well as a subtraction of the background from correlated $D$-meson
decays in case of dileptons.

% ____________________________________________________________________
\section*{Acknowledgments}
The authors are thankful to H. St\"ocker for suggesting this study
and his continuous interest. Furthermore, they would like to thank
M. Gorenstein and B. K\"ampfer for valuable discussions and
constructive proposals. Last but not least the authors acknowledge
the continuous discussions with J. Aichelin, H. Berrehrah, R.Marty, A.
Palmese, E. Seifert and T. Steinert. This work in part was supported
by the  LOEWE center HIC for FAIR as well as by BMBF.
P.M. acknowledges the financial support from the graduate school HGS-HIRe for FAIR.
The computational resources have been provided by the LOEWE-CSC.

\end{document}